\documentclass[twoside]{dis09}
\usepackage[latin1]{inputenc}
\usepackage[dvips]{graphicx,epsfig,color}
\usepackage{wrapfig,rotating}
\usepackage{cite,amssymb,amsmath,array,slashed}

\begin{document}

\title{
\vspace*{-2mm}\noindent
\hspace*{-4mm}
{\tiny DESY 09-125 \hfill SFB/CPP-09-77}\\
2- and 3-Loop Heavy Flavor Corrections to Transversity}

%*****************************1******************************************
% AUTHORS INFORMATION AREA
%***********************************************************************
\author{Johannes Bl\"umlein$^1$, Sebastian Klein$^1$ and Beat T\"odtli$^1$
%
% Optional short acknowledgment: remove next line if non-needed
%\thanks{This is an optional funding source acknowledgment.}
%
% DO NOT MODIFY THE FOLLOWING '\vspace' ARGUMENT
\vspace{.3cm}\\
%
% Addresses and institutions (remove "1- " in case of a single institution)
1- Deutsches Elektronen-Synchrotron DESY, Zeuthen\\
Platanenallee 6 - 15738 Zeuthen, Germany
%
% Remove the next three lines in case of a single institution
}
%***********************************************************************
% END OF AUTHORS INFORMATION AREA
%***********************************************************************

\maketitle

\begin{abstract}
We calculate the two- and three-loop massive operator matrix elements (OMEs) 
contributing to the heavy flavor Wilson coefficients of transversity. We obtain 
the complete result for the two-loop OMEs and compute the first thirteen Mellin
moments at three-loop order. As a by-product of the calculation, the moments 
$N=1$ to $13$ of the complete two-loop and the $T_F$-part of the three-loop 
transversity anomalous dimension are obtained. 
\end{abstract}

%--------------------------------------------------------------------------------
\section{Framework}
%--------------------------------------------------------------------------------

\vspace*{1mm}\noindent
The transversity distribution belongs to the three twist-2 parton distribution
functions (PDFs), together with those for unpolarized and polarized 
deep-inelastic scattering. It 
is a flavor non-singlet, chiral-odd distribution and can be measured in 
semi-inclusive deep-inelastic scattering (SIDIS) and via the polarized 
Drell-Yan 
process.~\footnote{For a review see Ref.~\cite{REV}.} Different experiments 
perform transversity measurements at the moment, cf. Refs.~\cite{EXP}. Recently, 
a first phenomenological parameterization has been given for the transversity up- 
and down-quark distributions in Ref.~\cite{PHEN}, the moments of which are in 
qualitative agreement with first lattice calculations \cite{LATT}. 

For semi-inclusive deeply inelastic charged lepton-nucleon scattering
$l N \rightarrow l' h + X$ the scattering cross section is given by 
%-----------------------------------------------------------------------
\begin{eqnarray}
\label{sidis1}
\frac{d^3 \sigma^{\rm SIDIS}}{dx dy dz} &=& \frac{4 \pi \alpha_{\mathrm{em}}^2 
s}{Q^4} 
\sum_{a =q,\overline{q}} e_a^2 x \Biggl\{\frac{1}{2} \left[1+(1-y)^2\right]
F_a(x,Q^2) D_a(z,Q^2) \nonumber\\ &&
- (1-y) |{\bf S}_\perp||{\bf S}_{h\perp}| \cos\left(\phi_S + \phi_{S_h}\right)
\Delta_T F_a(x,Q^2) \Delta_T D_a(z,Q^2)\Biggr\}~,
\end{eqnarray}
%-----------------------------------------------------------------------
after the ${\bf P}_{h \perp}$-integration has been performed,~\cite{REV}. 
We consider, for 
definiteness, only scattering cross sections free of ${\bf k}_\perp$- effects 
to refer to twist--2 quantities. $x$ and $y$ denote the Bjorken variables, $z$ 
the fragmentation 
variable, $Q^2 = - q^2$ the space-like 4--momentum transfer, $\alpha_{\mathrm{em}}$
the fine structure constant, $e_a$ the quark charge, and  $s$ the cms energy 
squared. 
${\bf S}_\perp$ and ${\bf S}_{h\perp}$ are the transverse spin vectors of the 
incoming nucleon $N$ and the measured hadron $h$. 
$F_a(z,Q^2), \Delta_T F_a(z,Q^2)$ and  
$D_a(z,Q^2), \Delta_T D_a(z,Q^2)$ 
denote the unpolarized and transversity structure- and fragmentation 
functions, respectively.
The angles $\phi_{S,S_h}$ are measured in the plane transverse to the 
$\gamma^* N$ axis between the $x$-axis and the respective vector.
In process (\ref{sidis1}) the spin of the \emph{transversely} polarized hadron 
$h$ has to be  measured.

The transversity distribution may also be measured in the transversely 
polarized Drell-Yan processes. 
In Mellin space the scattering cross section is given 
by, \cite{Vogelsang:1997ak},
%-----------------------------------------------------------------------
\begin{eqnarray}
\label{DY3}
\frac{d\Delta_T\sigma^{\rm DY}}{d \phi} &=& \frac{\alpha_{\rm em}^2}{9 s} 
\cos(2\phi)
\Delta_T H(N,M^2) \cdot \Delta_T C_q^{\rm DY}(N,M^2)
\end{eqnarray}
%-----------------------------------------------------------------------
where $N$ denotes the Mellin variable and $\phi$ is the azimuthal angle of one 
of the final state leptons $l^\pm$ relative to the axis 
defined by the transverse polarizations.
%-----------------------------------------------------------------------
\begin{equation*}
\Delta_T H(N,Q^2) = \sum_q e_q^2 \left[
                      \Delta_T q_1(N,Q^2)
                      \Delta_T \overline{q}_2(N,Q^2)
                  +   \Delta_T \overline{q}_1(N,Q^2)
                      \Delta_T q_2(N,Q^2)\right]
\end{equation*}
%-----------------------------------------------------------------------
 is a combination of transversity parton distributions for the incoming light 
(anti-)quarks, and $\Delta_T C^\mathrm{DY}_q(N,M^2)$ denotes the Wilson 
coefficient of 
the Drell-Yan process, with $M^2$ the invariant mass of the produced lepton 
pair.

Like in the case of unpolarized and polarized deep-inelastic processes 
transversity receives
heavy flavor corrections in higher orders in QCD. These are given by the 
corresponding heavy flavor Wilson coefficients. As for other  non-singlet 
quantities \cite{BUZA1,WHQ2}, these corrections start at $O(a_s^2)$, with
$a_s = \alpha_s/(4\pi)$.
In SIDIS one can tag $Q\bar{Q}$-production in the same way as in the 
deep-inelastic process, \cite{ABMK}. A measurement is possible in high 
luminosity experiments. In the Drell-Yan process, on the other hand, heavy 
flavor contributions emerge inclusively since there the final-state 
$l^+l^-$-pairs are measured in the first place. The calculation of the heavy 
quark Wilson coefficients for $Q^2 \gg m^2$ proceeds in the same way as in
unpolarized and polarized deep-inelastic 
scattering~\cite{BUZA1,WHQ2,HEAVY,BBK3} 
 
The complete Wilson coefficients for transversity can be decomposed into a 
light-
and a heavy quark contribution 
%--------------------------------------------------------------------------------
\begin{eqnarray}
\label{eqTR1}
{C}_q^\mathrm{TR}\left(x,\frac{Q^2}{\mu^2}, \frac{m^2}{\mu^2} \right)
= C_q^\mathrm{TR,light}\left(x,\frac{Q^2}{\mu^2}\right)
+H^\mathrm{TR}_q\left(x,\frac{Q^2}{\mu^2},\frac{m^2}{\mu^2}\right)\:.
\end{eqnarray}
%--------------------------------------------------------------------------------
As shown in \cite{BUZA1}, the heavy quark Wilson coefficient for hard 
processes factorizes into the light quark Wilson coefficients 
and  the massive operator matrix element $A_{qq,Q}^{\rm TR}$ at large 
enough scales $Q^2 \gg m^2$. We apply this to the heavy flavor Wilson 
coefficient for
transversity $H_q^{\rm TR}$ 
%------------------------------------------------------------------------
\small{\begin{eqnarray}
\label{HFAC}
H_{q}^\mathrm{TR} \left(x,\frac{Q^2}{\mu^2}, \frac{m^2}{\mu^2}\right) &=& 
C_{q}^\mathrm{TR,light} \left(x,\frac{Q^2}{\mu^2}\right) 
\otimes A_{qq,Q}^{\rm TR} 
\left(x,\frac{m^2}{\mu^2}\right)
\nonumber\\ 
&=&
 a_s^2\left[\Delta_T A_{qq,Q}^{(2),\rm NS,TR}(N_f+1) + \Delta_T
 \hat{C}_q^{(2)}(N_f)\right]
+  a_s^3\left[
\Delta_T A_{qq,Q}^{(3),\rm NS,TR}(N_f+1)\right. 
\nonumber\\ 
&+& \left.\Delta_T A_{qq,Q}^{(2),\rm NS,TR}(N_f+1) 
 \otimes \Delta_T 
{C}_q^{(1)}(N_f+1) + \hat{C}_q^{(3)}(N_f)\right]~.
\end{eqnarray}}
\normalsize
%--------------------------------------------------------------------------------

\vspace*{-5mm}
The aim of this article is to present a computation of the renormalized 
two- and three-loop heavy-flavor operator matrix elements contributing to 
transversity. Details of the calculation are given in Ref.~\cite{BKT1}.
The operator matrix element $\langle q\bigl|O^{\rm NS,TR}\bigr|q\rangle$ is 
given by a two-point Green's function containing a closed loop of a heavy 
quark $Q$ and external massless quarks $q$. The local operator is given by  
%--------------------------------------------------------------------------------
\begin{equation}\label{eq:operators}
O^{\mathrm{NS,TR}}_{F,a;\mu\mu_1 \ldots \mu_n} = i^{n} {\bf S} 
\Bigl[\overline{\psi}
\gamma_5\sigma_{\mu\mu_1} D_{\mu_2} \ldots D_{\mu_n} \frac{\lambda_a}{2}\psi\Bigr] 
- {\rm trace~terms}~,
\end{equation}
%--------------------------------------------------------------------------------
cf.~\cite{Geyer:1977gv}.
Here ${\bf S}$ denotes symmetrization of the Lorentz indices,  
$\sigma_{\mu\nu} = (i/2)\left[\gamma_\mu \gamma_\nu - \gamma_\nu 
\gamma_\mu\right]$, and $D_\mu$ is the covariant derivative.  The Green's 
function in $D = 4 + \varepsilon$ dimensions 
obeys the following vector decomposition 
%--------------------------------------------------------------------------------
\begin{multline}
 \hat{G}^{ij,\rm TR,NS}_{\mu,q,Q}=J_N\langle q|
O^{\mathrm{\rm NS,TR}}_{F,a;\mu\mu_1 \ldots \mu_n} |q\rangle
=\delta_{ij} \biggl\{\Delta_\rho\sigma^{\mu\rho}
\hat{\hat{A}}_{qq,Q}^{\rm
TR,NS}\left(\frac{\hat{m}^2}{\mu^2},\varepsilon,N\right)
+c_1 \Delta^\mu\bigr.\\+c_2 p^\mu
+c_3\gamma^\mu\slashed{p}+c_4\slashed{\Delta}\slashed{p}\Delta^\mu
+c_5\slashed{\Delta}\slashed{p}p^\mu\biggr\} \left(\Delta\cdot p\right)^{N-1}
\end{multline}
%--------------------------------------------------------------------------------
contracting the OME with a 
source term $J_N=\Delta^{\mu_1}\ldots\Delta^{\mu_N}$, with 
$\Delta^2=0$, with $p$ the parton momentum.
It determines the un-renormalized massive OME 
%--------------------------------------------------------------------------------
\begin{multline}
\hat{\hat{A}}_{qq,Q}^{\rm TR,NS}\left(\frac{\hat{m}^2}{\mu^2},\varepsilon,N\right)
=
\frac{-i\:\delta^{ij}}{4N_c\left(\Delta\cdot p\right)^{N+1}\left(D-2\right)}
\biggl\{\mathrm{Tr}\left[\slashed{\Delta}\slashed{p}p^\mu 
\hat{G}^{ij,\rm TR,NS}_{\mu,q,Q}\right]\\ 
-\Delta\cdot p\:\mathrm{Tr}
\left[p^\mu \hat{G}^{ij,\rm TR,NS}_{\mu,q,Q}\right]
+i \Delta\cdot p\:\mathrm{Tr}\left[p^\mu 
\hat{G}^{ij,\rm TR,NS}_{\mu,q,Q}\right] 
\biggr\}\: .
\end{multline}
%--------------------------------------------------------------------------------
A total of 129 diagrams contribute, which were generated using {\tt QGRAF} 
\cite{Nogueira:1991ex}. These were 
projected onto $\hat{\hat{A}}_{qq,Q}^{TR,NS}$, cf.~\cite{BBK3}, using 
codes written in {\tt FORM}~\cite{Vermaseren:2000nd}. 
After tensor reduction, the loop integrals are of the tadpole-type, since 
the single external quark is on-shell and massless. The integrals were then 
evaluated using {\tt MATAD} \cite{Steinhauser:2000ry}. The renormalization of 
the OMEs is described in Ref.~\cite{BBK3}.

After mass- and charge renormalization one obtains the massive OMEs
in the on-mass-shell scheme, cf. \cite{BBK3},  
%--------------------------------------------------------------------------------
\begin{eqnarray}
\Delta_T A_{qq,Q}^{(2),{\rm NS, \overline{\rm MS}}}&=&
                   \frac{\beta_{0,Q}\gamma_{qq}^{(0),\rm TR}}{4}
                     \ln^2 \Bigl(\frac{m^2}{\mu^2}\Bigr)
                  +\frac{\hat{\gamma}_{qq}^{(1), {\rm TR}}}{2}
                     \ln \Bigl(\frac{m^2}{\mu^2}\Bigr)
                  +a_{qq,Q}^{(2),{\rm TR}}
                  -\frac{\beta_{0,Q}\gamma_{qq}^{(0),\rm TR}}{4}\zeta_2~, 
\nonumber \\
                   \label{Aqq2NSTRQMSren} \\
%%%
     \Delta_T    A_{qq,Q}^{(3),{\rm NS}, \overline{\rm MS}}&=&
      -\frac{\gamma_{qq}^{(0),{\rm TR}}\beta_{0,Q}}{6}
           \Bigl(
                  \beta_0
                 +2\beta_{0,Q}
           \Bigr)
              \ln^3 \Bigl(\frac{m^2}{\mu^2}\Bigr)
          +\frac{1}{4}
           \Biggl\{
                    2\gamma_{qq}^{(1),{\rm TR}}\beta_{0,Q}\nonumber
\\
&&
                   -2\hat{\gamma}_{qq}^{(1),{\rm TR}}
                              \Bigl(
                                     \beta_0
                                    +\beta_{0,Q}
                              \Bigr)
                   +\beta_{1,Q}\gamma_{qq}^{(0),{\rm TR}}
           \Biggr\}
              \ln^2 \Bigl(\frac{m^2}{\mu^2}\Bigr)
          +\frac{1}{2}
           \Biggl\{
                    \hat{\gamma}_{qq}^{(2),{\rm TR}}
\nonumber\\ &&
                   -\Bigl(
                            4a_{qq,Q}^{(2),{\rm TR}}
                           -\zeta_2\beta_{0,Q}\gamma_{qq}^{(0),{\rm TR}}
                                     \Bigr)(\beta_0+\beta_{0,Q})
                   +\gamma_{qq}^{(0),{\rm TR}}\beta_{1,Q}^{(1)}
           \Biggr\}
              \ln \Bigl(\frac{m^2}{\mu^2}\Bigr)
\nonumber\\&&
          +4\overline{a}_{qq,Q}^{(2),{\rm TR}}(\beta_0+\beta_{0,Q})
          -\gamma_{qq}^{(0)}\beta_{1,Q}^{(2)}
          -\frac{\gamma_{qq}^{(0),{\rm TR}}\beta_0\beta_{0,Q}\zeta_3}{6}
          -\frac{\gamma_{qq}^{(1),{\rm TR}}\beta_{0,Q}\zeta_2}{4}
\nonumber\\ \nonumber \\&&
          +2 \delta m_1^{(1)} \beta_{0,Q} \gamma_{qq}^{(0),{\rm TR}}
          +\delta m_1^{(0)} \hat{\gamma}_{qq}^{(1),{\rm TR}}
          +2 \delta m_1^{(-1)} a_{qq,Q}^{(2),{\rm TR}}
          +a_{qq,Q}^{(3),{\rm TR}}~, \label{Aqq3NSTRQMSren}
\label{AT3}
\end{eqnarray}
%--------------------------------------------------------------------------------
at 2-- and 3--loops. Here, $\zeta_k$ denotes the Riemann $\zeta$-function at 
$\gamma_{qq}^{(l), \rm TR}$ are the transversity anomalous 
dimensions for $l = 0,1,2$ in LO \cite{gLO}, NLO \cite{Vogelsang:1997ak,NLOg}, 
and NNLO \cite{GRAC}, with $\hat{f}(N_f) = f(N_f+1) - f(N_f)$.
For the other quantities we refer to \cite{BBK3}. 
The new terms being  calculated are
$a_{qq,Q}^{(2),{\rm TR}}(N),~\overline{a}_{qq,Q}^{(2),{\rm TR}}(N)$ and
$a_{qq,Q}^{(3),{\rm TR}}(N)$, and for the higher values of $N$, 
$\hat{\gamma}_{qq}^{(2), \rm TR}(N)$.

\section{Results}
\subsection{Massive Operator Matrix Elements}

\vspace*{1mm}\noindent
At $O(a_s^2)$ the  massive operator matrix elements for transversity 
$\Delta_T A_{qq,Q}^{(2),{\rm NS, \overline{\rm MS}}}$ are obtained for general 
values of $N$, cf. Eq.~(\ref{Aqq2NSTRQMSren}). The un-renormalized OME is 
computed
to $O(\varepsilon)$ to also extract the functions $\overline{a}_{qq,Q}^{\rm 
TR, (2)}(N)$. The new terms at 2--loops are
$a_{qq,Q}^{\mathrm{TR},(2)}$ and $\bar{a}_{qq,Q}^{\mathrm{TR},(2)}$,
cf. Eqs.~(\ref{Aqq2NSTRQMSren}, \ref{Aqq3NSTRQMSren}):
%--------------------------------------------------------------------------------
\begin{multline}
a_{qq,Q}^{\rm TR, (2)}(N) = C_F T_F \Biggl\{ 
-\frac{8}{3}    S_3
+\frac{40}{9}   S_2
-\left[\frac{224}{27} 
      +\frac{8}{3}\zeta_2 \right] S_1
%\\
+2 \zeta_2
+
{\frac{ \left( 24+73\,N+73\,{N}^{2} \right)}{18 N \left( 1+N \right) }}
\Biggr\} 
\end{multline}
%--------------------------------------------------------------------------------
%--------------------------------------------------------------------------------
\begin{multline}
 \overline{a}_{qq,Q}^{\rm TR, (2)}(N) = C_F T_F \Biggl\{
- \left[
{\frac {656}{81}}\, 
+{\frac {20}{9}}\, \zeta_2
+{\frac {8}{9}}\, \zeta_3 \right] S_1
+\left[{\frac {112}{27}}\, +\frac{4}{3}\, \zeta_2 \right] S_2
-{\frac {20}{9}}\, S_3
\\ 
+\frac{4}{3}\, S_4
+\frac{1}{6}\, \zeta_2
+\frac{2}{3}\, \zeta_3
+{\frac { 
\left( -144-48\,N+757\,{N}^{2}+1034\,{N}^{3}+517\,{N}^{4} \right) }
{216 {N}^{2} \left( 1+N \right) ^{2}}} \Biggr\}~,
\end{multline}
%--------------------------------------------------------------------------------
with $S_k \equiv S_k(N)$ the single harmonic sums.

At $O(a_s^3)$ the moments $N=1$ to 13 were computed for the massive OMEs, as 
e.g. 
%-----------------------------------------------------------------------
\begin{eqnarray}
    \Delta_T~A_{qq,Q}^{(3),{\rm NS}, \overline{\rm MS}}(13)&=&C_FT_F\Biggl\{
        \Bigl(
               \frac{1751446}{110565}C_A
              -\frac{7005784}{1216215}T_F(N_f+2)
        \Bigr)\ln^3\Bigl(\frac{m^2}{\mu^2}\Bigr)
\nonumber\\&&\hspace{-20mm}
       +\Bigl(
              -\frac{20032048197492631}{2193567563187000}C_F
              -\frac{137401473299}{8027019000}C_A
              -\frac{93611152819}{3652293645}T_F
        \Bigr)\ln^2\Bigl(\frac{m^2}{\mu^2}\Bigr)
\nonumber\\&&\hspace{-20mm}
       +\Bigl[
             \Bigl(
                    \frac{1705832327329042449983}{263491335690022440000}
                   +\frac{7005784}{45045}\zeta_3
             \Bigr)C_F
            +\Bigl(
                    \frac{3385454488248191237}{65807026895610000}
\nonumber\\&&\hspace{-20mm}
                   -\frac{7005784}{45045}\zeta_3
             \Bigr)C_A
              -\frac{458114791076413771}{6580702689561000}N_fT_F
              -\frac{217179304}{3648645}T_F
        \Bigr]\ln\Bigl(\frac{m^2}{\mu^2}\Bigr)
\nonumber\\&&\hspace{-20mm}
      +\Bigl(
             -\frac{7005784}{135135}B_4
             +\frac{3502892}{15015}\zeta_4
             -\frac{81735983092}{243486243}\zeta_3
\nonumber\\&&\hspace{-20mm}
             +\frac{55376278299522733837425052493}{122080805651901196900800000}
       \Bigr)C_F
      +\Bigl(
              \frac{3502892}{135135}B_4
             -\frac{3502892}{15015}\zeta_4
\nonumber\\&&\hspace{-20mm}
             +\frac{4061479439}{12162150}\zeta_3
             -\frac{3486896974743882556775647}{12935029206601101600000}
       \Bigr)C_A
\nonumber\\&&\hspace{-20mm}
      +\Bigl(
             -\frac{279922752632160355860697}{3557133031815302940000}
             +\frac{56046272}{1216215}\zeta_3
       \Bigr)T_FN_f
\nonumber\\&&\hspace{-20mm}
      +\Bigl(        \frac{291526550302760070155303}{7114266063630605880000}
             -\frac{14011568}{173745}\zeta_3
       \Bigr)T_F
                    \Biggr\}~,
\end{eqnarray}
%-----------------------------------------------------------------------
where 
%-----------------------------------------------------------------------
$$B_4=-4\zeta_2\ln^2\left(2\right)+\frac{2}{3}\ln^4(2)-\frac{13}{2}\zeta_4
+16\mathrm{Li}_4\left(\frac{1}{2}\right)\:.$$
%-----------------------------------------------------------------------
Like for the massive OMEs in case of unpolarized deep-inelastic scattering, 
the structure of
$\Delta_T~A_{qq,Q}^{(3),{\rm NS}, \overline{\rm MS}}(N)$ is widely known
for general values of $N$, except the contributions due to
the finite part $a_{qq,Q}^{(3),{NS}}$  and the 3-loop anomalous dimension 
$\hat{\gamma}_{qq}^{(2), \rm TR}(N)$. One notices the
cancellation of all $\zeta_2$ terms in 
$\Delta_T~A_{qq,Q}^{(3),{\rm NS}, \overline{\rm MS}}(N)$
after renormalization. 
%%%%%%%%%%%%%%%%%%%%%%%%%%%%%%%%%%%%%%%%%%%%%%%%%%%%%%%%%%%%%%%%%%%%%%%%%%%%%%%%%%%
\subsection{Anomalous Dimensions}
%%%%%%%%%%%%%%%%%%%%%%%%%%%%%%%%%%%%%%%%%%%%%%%%%%%%%%%%%%%%%%%%%%%%%%%%%%%%%%%%%%%

\vspace{1mm}\noindent
The transversity anomalous dimension is given by
%-----------------------------------------------------------------------
\begin{equation}
{\gamma}^{\rm TR}_{qq}\left(N,a_s\right)
=\sum_{i=1}^\infty 
a_s^i {\gamma}^{\left(i\right),\rm TR}_{qq}\left(N\right).
\end{equation}
%-----------------------------------------------------------------------
From Eq.~(\ref{AT3}) one may determine the complete 2-loop anomalous dimension
\cite{Vogelsang:1997ak,NLOg} and the $T_F$-part of the 3-loop anomalous
dimension \cite{GRAC}. We agree with the results given in 
\cite{Vogelsang:1997ak,NLOg} and
confirm the $T_F$-contributions for the moments $N=1$ to 8 given in 
Refs.~\cite{GRAC}. Furthermore, we obtain $\hat{\gamma}_{qq}^{(3), 
\mathrm{TR}} = 
{\gamma}_{qq}^{(3),\mathrm{TR}}({N_f+1})
-{\gamma}_{qq}^{(3),\mathrm{TR}}(N_f)$ newly for $N=9$ to 13, as e.g.
%-----------------------------------------------------------------------
\begin{equation}
\begin{split}
\hat{\gamma}_{qq}^{(3), \mathrm{TR}} \left(N=13\right)=& -C_F T_F \Biggl[
{\frac {36713319015407141570017}{131745667845011220000}}\,{C_F}
-{\frac {14011568}{45045}}\, (C_F - C_A) \zeta_3
\\& 
+{\frac {66409807459266571}{3290351344780500}}\,{T_F} (1+2 N_f)
+{\frac {6571493644375020121}{65807026895610000}}\,{C_A}\nonumber
\Biggr]\:.
\end{split}
\end{equation}
%-----------------------------------------------------------------------
%%%%%%%%%%%%%%%%%%%%%%%%%%%%%%%%%%%%%%%%%%%%%%%%%%%%%%%%%%%%%%%%%%%%%%%%%%%%%%%%%%%
\subsection{A Remark on the Soffer Bound}
%%%%%%%%%%%%%%%%%%%%%%%%%%%%%%%%%%%%%%%%%%%%%%%%%%%%%%%%%%%%%%%%%%%%%%%%%%%%%%%%%%%

\vspace{1mm}\noindent
If the Soffer inequality \cite{Soffer:1994ww}
%-----------------------------------------------------------------------
\begin{eqnarray}
|\Delta_T f(x,Q^2)|
\leq \frac{1}{2} \left[ f(x,Q^2) + \Delta f(x,Q^2)\right]
\label{SOF1}
\end{eqnarray}
%-----------------------------------------------------------------------
holds for the non-perturbative PDFs in Eq.~(\ref{SOF1})
one may check its generalization from  $f_i \rightarrow F_i$ for the 
corresponding structure functions. This includes the non-singlet evolution
operator (Eq.~(6), Ref.~\cite{BBG}) and the heavy flavor Wilson coefficient. 
At 
perturbative scales, it holds for the evolution operator 
\cite{BKT1}, generalizing a result from \cite{Vogelsang:1997ak} for the 
moments $N=1$ to 13 at 3--loops. For the heavy 
quark Wilson coefficients in SIDIS we only know the massive OMEs so far. 
As shown in Ref.~\cite{BKT1}, a final conclusion can only be drawn knowing 
the yet undetermined massless Wilson coefficients. Here the  
difference $[A_{qq,Q}^{\rm V} - A_{qq,Q}^{\rm TR}](x)$ of the massive OMEs, 
shows a 
sign 
change to negative values for  $Q^2/m^2$ in the physical range. 
For large scales $Q^2/m^2 \gg 1$ positive values are obtained.
%------------------------------------------------------------------------------
%------------------------------------------------------------------------

% ****************************************************************************
\end{document}